\documentclass[journal]{IEEEtran}
\usepackage{amsmath,amsfonts}
\usepackage{array}
\usepackage[caption=false,font=footnotesize,labelfont=rm,textfont=rm]{subfig}
\usepackage{textcomp}
\usepackage{stfloats}
\usepackage{url}
\usepackage{verbatim}
\usepackage{graphicx}
\usepackage{cite}
\usepackage{epstopdf}
\usepackage{tabularx}
\newcommand{\tabincell}[2]{\begin{tabular}{@{}#1@{}}#2\end{tabular}} 
\usepackage{xcolor}
\usepackage{bm} 

\usepackage{booktabs}
\usepackage{multirow}
\usepackage{siunitx}  
\usepackage{threeparttable} 

\usepackage{algorithm}
\usepackage{algpseudocode}

\def\BibTeX{{\rm B\kern-.05em{\sc i\kern-.025em b}\kern-.08em
		T\kern-.1667em\lower.7ex\hbox{E}\kern-.125emX}}

\begin{document}
	
\ifdefined \GramaCheck
\newcommand{\CheckRmv}[1]{}
\newcommand{\figref}[1]{Figure 1}%
\newcommand{\tabref}[1]{Table 1}%
\newcommand{\secref}[1]{Section 1}
\newcommand{\algref}[1]{Algorithm 1}
\renewcommand{\eqref}[1]{Equation 1}
\else
\newcommand{\CheckRmv}[1]{#1}
\newcommand{\figref}[1]{Fig.~\ref{#1}}%
\newcommand{\tabref}[1]{Table~\ref{#1}}%
\newcommand{\secref}[1]{Sec.~\ref{#1}}
\newcommand{\AlgRef}[1]{Algorithm~\ref{#1}}
\renewcommand{\eqref}[1]{(\ref{#1})}
\fi

\title{Semantic Satellite Communications for Synchronized Audiovisual Reconstruction}

\author{
	Fangyu Liu,~\IEEEmembership{Graduate Student Member,~IEEE,}
	~Peiwen Jiang,~\IEEEmembership{Member,~IEEE,}
	
	~Wenjin Wang,~\IEEEmembership{Member,~IEEE,}
	~Xiao Li,~\IEEEmembership{Member,~IEEE,}
	and Shi Jin,~\IEEEmembership{Fellow,~IEEE}
	
	\thanks{Fangyu Liu, Peiwen Jiang, Wenjin Wang, Xiao Li, and Shi Jin are with the School of Information Science and Engineering, Southeast University, Nanjing 210096, China (e-mail: fangyuliu@seu.edu.cn; peiwenjiang@seu.edu.cn; wangwj@seu.edu.cn; li\_xiao@seu.edu.cn; jinshi@seu.edu.cn).}
}
\maketitle

\begin{abstract}
Satellite communications face severe bottlenecks in supporting high-fidelity synchronized audiovisual services, as conventional schemes struggle with cross-modal coherence under fluctuating channel conditions, limited bandwidth, and long propagation delays. To address these limitations, this paper proposes an adaptive multimodal semantic transmission system tailored for satellite scenarios, aiming for high-quality synchronized audiovisual reconstruction under bandwidth constraints. Unlike static schemes with fixed modal priorities, our framework features a dual-stream generative architecture that flexibly switches between video-driven audio generation and audio-driven video generation. By dynamically decoupling semantics, the system transmits the priority modality and recovers the other via cross-modal generation. To balance reconstruction quality and transmission overhead, a dynamic keyframe update mechanism adaptively maintains the shared knowledge base according to wireless scenarios and user requirements. Furthermore, a large language model based decision module is introduced to enhance system adaptability. By integrating satellite-specific knowledge, this module jointly considers task requirements and channel factors such as weather-induced fading to proactively adjust transmission paths and generation workflows. Simulation results demonstrate that the proposed system significantly reduces bandwidth consumption while achieving high-fidelity audiovisual synchronization, improving transmission efficiency and robustness in challenging satellite scenarios.
\end{abstract}


\section{Introduction}
\IEEEPARstart{W}{ith} the growing demand for global connectivity, satellite communications have become indispensable for maritime, aerial, and disaster-relief operations where terrestrial infrastructure is unavailable \cite{8700142}. However, links between ground stations and diverse user terminals are subject to severe physical layer constraints, including rain attenuation, significant Doppler shifts in non-geostationary orbits, and propagation delays exceeding several hundred milliseconds \cite{9079470}. Although conventional techniques such as adaptive modulation \cite{8638565} and robust beamforming \cite{9134413} can improve spectral efficiency, they struggle to support data-intensive multimodal streams, for example synchronized audiovisual content, when transponder rates are limited to the kbps level. As a result, unpredictable satellite channel impairments remain a fundamental bottleneck for high-fidelity multimedia services.

Semantic communication has recently emerged as a promising paradigm to mitigate bandwidth limitations in satellite networks \cite{qin2021semantic}. Unlike traditional communication systems, semantic communication improves transmission efficiency by extracting and transmitting only task-relevant semantics. Knowledge bases play a central role in this process by enabling context-aware transmission, enhancing inference capability, and substantially reducing data traffic \cite{10628028,10554663}. In satellite communications, a shared knowledge base between the transmitter and receiver enables adaptive semantic strategies, thereby maintaining transmission quality and efficiency under highly constrained channel conditions.

Prior studies have successfully applied semantic communication to text \cite{xie2021lite} and images \cite{8723589}. However, video transmission remains a bottleneck due to its data volume and temporal complexity. Efficiently transmitting such bandwidth-intensive content is now a priority in the development of next-generation networks \cite{9837870,9953110,10283754,9955991}. Existing semantic video transmission methods can be broadly classified into joint source-channel coding (JSCC) and generative model based approaches. As an early JSCC-based solution, DeepWiVe \cite{9837870} jointly encodes and decodes multiple frames to exploit temporal correlations and improve transmission quality. Similarly, the DVST system \cite{9953110} adopts nonlinear transformations and conditional encoding to extract inter-frame semantic features, achieving superior performance over conventional schemes. To further enhance efficiency, VISTA \cite{10283754} separates video content into dynamic and static components for independent encoding. In contrast, generative model based approaches such as SVC \cite{9955991} utilize pre-shared static frames as a knowledge base and transmit only key semantic features associated with content variations, from which the receiver generates the full video content, thereby substantially reducing transmission overhead with limited loss of visual fidelity.

Despite these advances, most existing methods focus on single modality video transmission, whereas practical applications often require synchronized audiovisual data, thereby demanding higher adaptability. Recent studies have explored cross-modal correlations for semantic encoding, including image-text transmission for visual question answering \cite{9830752} and multimodal fusion for audiovisual event localization \cite{yu2024pilot}. In video conferencing scenarios, several frameworks transmit 3D morphable model (3DMM) parameters to reconstruct facial videos and assist audio generation \cite{10872781}, or use text and audio to drive synchronized video synthesis \cite{10740049,9953071,jiang2024large}. 
However, modality prioritization and the selection of cross-modal generation paths in these methods are fixed at the design stage. This rigid configuration prevents the system from flexibly adjusting modality importance based on task requirements, such as prioritizing audio in emergency services. Moreover, existing knowledge base driven generative semantic systems \cite{9955991,10872781} generally lack context awareness and dynamic update mechanisms. This prevents the system from balancing generation performance with bandwidth consumption, leading to resource waste or outdated information in constrained channels.
To address these limitations, the integration of generative foundation models, particularly large language models (LLMs) \cite{achiam2023gpt}, offers a promising direction by enabling human-like contextual understanding and adaptive semantic allocation.

Besides content adaptability, the inherent instability of satellite links requires robust channel-adaptive transmission strategies. Strong inter-modality dependence makes multimodal systems vulnerable to cascading errors under channel fading. 
To address this, traditional approaches rely on rule-based or lookup-table methods. However, the complex interplay between satellite dynamics and diverse task requirements leads to an exponentially growing state space. This makes it impossible to enumerate all scenario combinations, often resulting in suboptimal strategy selection or improper resource allocation.
To cope with such challenges, recent work has begun to exploit the reasoning and planning capabilities of LLMs to design active transmission strategies \cite{jiang2025semantic}. By dynamically sensing channel conditions and adjusting resource allocation, these approaches can mitigate the impact of fading. Incorporating foundation models into satellite semantic communication therefore enables adaptive optimization of both semantic compression and transmission reliability, ensuring stable end-to-end multimodal services under fluctuating channel conditions.

Motivated by the need for adaptability in uncertain environments, this paper proposes an adaptive semantic satellite transmission system governed by an LLM-based agent. Departing from traditional frameworks that passively execute fixed rules, this system marks a paradigm shift from static rule matching to intelligent planning. To enhance flexibility, a dual-stream multimodal generation module is employed, enabling dynamic switching between video-driven audio generation and audio-driven video generation. To balance reconstruction quality and bandwidth consumption, a dynamic keyframe update mechanism is introduced for knowledge base maintenance. Furthermore, serving as the core controller, the LLM-based agent coordinates these generation paths and update mechanisms. By understanding both task requirements and physical constraints, the agent addresses key operational challenges that traditional methods cannot, ensuring high-fidelity synchronized reconstruction even in complex satellite scenarios. The main contributions of this work are summarized as follows.

\begin{itemize}
\item \textbf{Adaptive multimodal synchronization via dual-stream generative strategies:} The proposed system dynamically prioritizes key modality transmission by switching between video-driven and audio-driven generation according to task requirements. This ensures high-fidelity audiovisual reconstruction and robust synchronization under severe satellite bandwidth constraints and channel fading.

\item \textbf{Dynamic knowledge base management for bandwidth-quality trade-off:} A dynamic keyframe update mechanism is introduced to mitigate the impact of outdated shared knowledge bases. By adapting the update frequency to wireless conditions and user requirements, the system achieves an effective balance between reconstruction quality and bandwidth consumption in resource-constrained satellite environments.

\item \textbf{Intelligent scenario- and task-aware adaptivity:} We design an LLM-driven decision framework for satellite multimodal semantic communication. By jointly considering task requirements, user preferences, and satellite link environment, the agent performs workflow selection, transmission strategy adaptation, and resource configuration, thereby improving transmission efficiency and reconstruction quality under dynamic channel constraints.
\end{itemize}

The remainder of this paper is organized as follows. Section \ref{section:System Model} describes the system model. Section \ref{section:Proposed Adaptive Audiovisual Synchronous Semantic Transmission System} presents the proposed framework. Section \ref{section:Numerical Results} provides experimental results. Finally, Section \ref{section:Conclusion} concludes the paper.

\section{System Model} \label{section:System Model} 
In this section, we first describe the semantic satellite communication environment, followed by an overview of the multimodal semantic transmission framework.

\subsection{Semantic Satellite Communication Scenarios}

This paper considers a satellite-based multimodal transmission scenario, transmitting face videos and corresponding audio, as shown in \figref{satellite}. 
The ground transmitter performs task-oriented semantic encoding on the audio and video data and transmits them to the satellite. The satellite, acting as a relay, forwards the received semantic features through the downlink to the ground receiver for reconstruction.
The design of such a system is primarily challenged by the following factors.

\begin{itemize}
	\item \textbf{Dynamic and constrained channel:} The satellite link is affected by FSPL, rainfall attenuation (RA) \cite{series2015propagation}, and Doppler shifts caused by high-speed motion. To ensure reliable transmission, conservative modulation and coding schemes are required, which severely limit the effective bandwidth and throughput.
	
	\item \textbf{Diverse service requirements:} Satellite services support heterogeneous applications. For example, video conferencing prioritizes audiovisual synchronization, whereas disaster broadcasting emphasizes audio intelligibility. Such diversity necessitates a task-aware adaptive transmission strategy under limited resources.
\end{itemize}

\CheckRmv{\begin{figure} [ht] 
		\centerline{\includegraphics[width=3.5in]{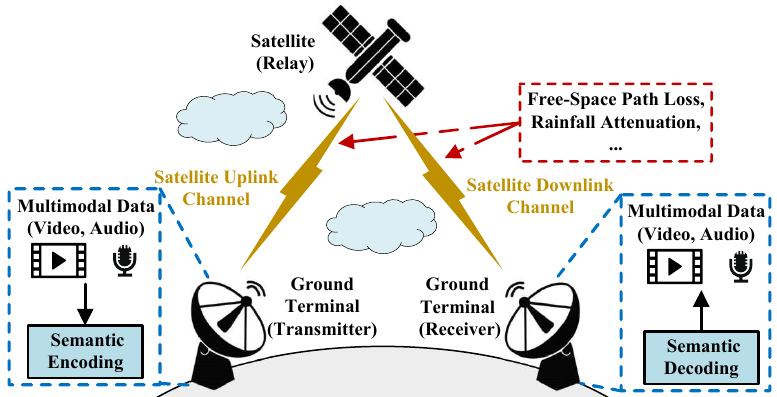}}
		\caption{System model of satellite-ground transmission, including satellite, transmission link, and ground terminal.}
		\label{satellite}
\end{figure}}

\subsection{Semantic Transmission Framework}
To transmit multimodal audiovisual data $\mathbf{M}$, $\mathbf{M}$ is first decomposed into video $\mathbf{V}$ and the corresponding audio $\mathbf{A}$. At the ground transmitter, semantic source encoding is applied to extract essential features, given by
\CheckRmv{
	\begin{align}
		\mathbf{S}_{\text{video}} &= S_{\text{video}}(\mathbf{V}),
		\label{eq1}   \\ 
		\mathbf{S}_{\text{audio}} &= S_{\text{audio}}(\mathbf{A}),
		\label{eq2}
	\end{align}
}
where $\mathbf{S}_{\text{video}}$ and $\mathbf{S}_{\text{audio}}$ denote the extracted semantic features, and $S_{\text{video}}(\cdot)$ and $S_{\text{audio}}(\cdot)$ represent the semantic source encoders. To preserve audiovisual synchronization, $\mathbf{S}_{\text{video}}$ and $\mathbf{S}_{\text{audio}}$ are multiplexed into a unified data stream and mapped onto OFDM symbols $\mathbf{X}$ with $N_t$ time symbols and $N_f$ subcarriers.

\CheckRmv{\begin{figure*}[!h]
		\centerline{\includegraphics[width=6.3in]{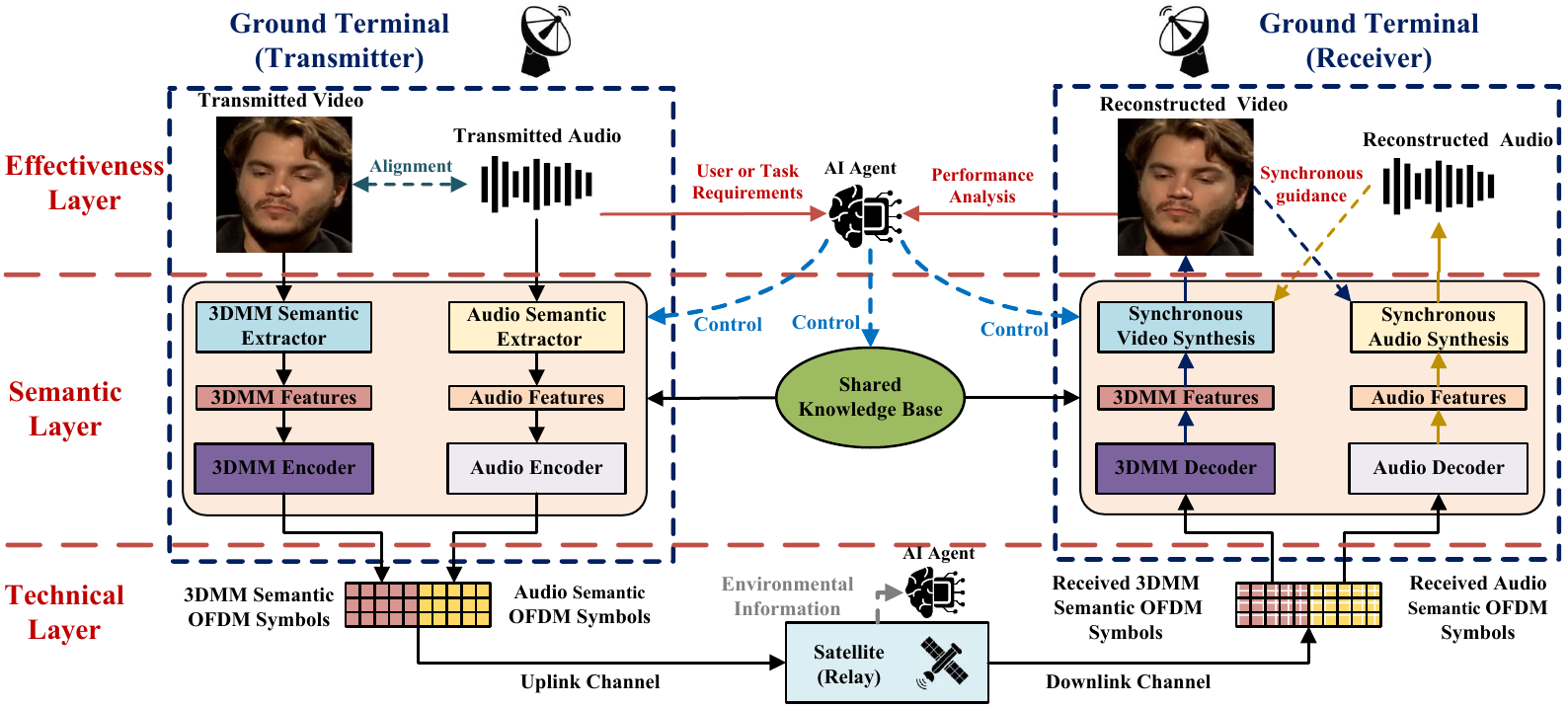}}
		\caption{Structure of the proposed adaptive audiovisual synchronous semantic transmission system for satellite communication scenarios.}
		\label{Proposed system}
\end{figure*}}

After transmission over the uplink and downlink channels, the received signal at the ground terminal is expressed as
\CheckRmv{ 
	\begin{equation}
		\mathbf{Y}=\mathbf{H}_\text{down}\odot\mathbf{H}_\text{up}\odot \mathbf{X}+\mathbf{Z},
		\label{eq3}
	\end{equation}
}
where $\odot$ denotes element-wise multiplication, $\mathbf{Z}$ represents additive noise, and $\mathbf{H}_\text{up} \in \mathbb{C}^{N_f \times N_t}$ and $\mathbf{H}_\text{down} \in \mathbb{C}^{N_f \times N_t}$ are the uplink and downlink channel frequency responses, respectively. The corresponding channel impulse response is modeled as
\CheckRmv{
	\begin{equation}
		h(t,\tau) = \sum_{l=0}^{L-1} h_l \delta(\tau-\tau_l)\exp(j2\pi f_l t),
		\label{eq4}
	\end{equation}
}
where $L$ is the number of channel taps, $h_l$ denotes the channel gain, $\tau_l$ is the propagation delay, and $f_l$ is the Doppler shift of the $l$th path.

Small-scale fading effects, such as Doppler shifts, can be mitigated by OFDM and dense pilot-aided channel estimation. However, large-scale fading caused by long propagation distance and weather conditions, including FSPL and RA, significantly degrades the received SNR. The FSPL \cite{9097410} is given by
\CheckRmv{
	\begin{equation}
		\text{FSPL(dB)} 
	    = 20\log_{10}(d) + 20\log_{10}(f) - 147.56,
		\label{eq5}
	\end{equation}	
}
where $f$ is the carrier frequency in hertz, and $d$ is the satellite link distance in meters, determined by the orbit altitude and the ground terminal elevation angle.
RA refers to signal degradation caused by propagation through rain regions. According to the ITU R P.618 model \cite{series2015propagation}, the attenuation is expressed as
\CheckRmv{\begin{equation}
		\text{RA(dB)} = \gamma_R \cdot L_e,
		\label{eq6}
\end{equation}}
where $\gamma_R$ is the specific attenuation determined by rainfall rate and carrier frequency, and $L_e$ is the effective path length through the rain region.

Accordingly, the SNR of the satellite link \cite{10223379} is given by
\CheckRmv{\begin{equation}
		\text{SNR(dB)} = P_\text{t} + G_\text{t} + G_\text{s} + G_\text{r} -\text{RA}_\text{all}-\text{FSPL}_\text{all} - N,
		\label{eq7}
\end{equation}} 
where $P_\text{t}$ is the transmit power in dBm, $G_\text{t}$ and $G_\text{r}$ are the transmit and receive antenna gains in dBi, $G_\text{s}$ is the equivalent gain of the relay satellite, $\text{FSPL}_\text{all}$ and $\text{RA}_\text{all}$ are the total FSPL and RA of the satellite channel, respectively, and $N$ denotes the receiver thermal noise power, computed as $N = 10\log_{10}(k{\cal T}B)$. Here, $k$ is the Boltzmann constant, $B$ is the system bandwidth, and ${\cal T}$ is the noise temperature.

At the receiver, an estimate $\widehat{\mathbf{H}}$ is obtained via least-squares channel estimation, and equalization yields
\CheckRmv{\begin{equation}
		\widehat{\mathbf{X}} = \mathbf{Y} \oslash \widehat{\mathbf{H}},
		\label{eq8}
\end{equation}}
where $\oslash$ denotes element-wise division. The noisy semantic features $\widehat{\mathbf{S}}_{\text{video}}$ and $\widehat{\mathbf{S}}_{\text{audio}}$ are extracted from $\widehat{\mathbf{X}}$ and fed into the semantic decoder for synchronized audiovisual reconstruction, given by
\CheckRmv{\begin{equation}
			\widehat{\mathbf{V}}, \widehat{\mathbf{A}} = S_{\text{sync}}\!\left(\widehat{\mathbf{S}}_{\text{video}}, \widehat{\mathbf{S}}_{\text{audio}}, \mathbf{KB}\right),
		\label{eq9}
\end{equation}}
where $S_{\text{sync}}(\cdot)$ denotes the synchronization-aware semantic decoder. Depending on task requirements, it adaptively selects either video-driven audio generation or audio-driven video generation. The semantic knowledge base $\mathbf{KB}$, similar to \cite{9955991,10872781}, contains the user reference image and supports identity-consistent video generation.

\section{Proposed Adaptive Audiovisual Synchronous Semantic Transmission System}
\label{section:Proposed Adaptive Audiovisual Synchronous Semantic Transmission System}
In this section, we first present the proposed system framework and describe the dual-stream audiovisual semantic coding process. We then analyze the impact of shared keyframes on generative performance, followed by a dynamic knowledge base updating mechanism designed for diverse application scenarios. Finally, an LLM-based decision module is introduced to adaptively optimize transmission paths and generation workflows under varying satellite channel conditions.

\subsection{System Framework}\label{section:Proposed}
As shown in \figref{Proposed system}, the proposed system consists of three core layers, namely the effectiveness layer, the semantic layer, and the technical layer, all integrated with a shared semantic knowledge base. These components jointly optimize transmission performance for resource-constrained satellite links.

\subsubsection{Effectiveness Layer}
The effectiveness layer focuses on semantic transmission quality and user experience by evaluating task-specific performance metrics, such as facial video reconstruction fidelity and audio semantic accuracy.

\subsubsection{Semantic Layer}
Guided by the LLM agent, the semantic transmitter selectively extracts task-relevant features from video $\mathbf{V}$ and audio $\mathbf{A}$. Unlike single-modal SVC \cite{9955991} or fixed-priority schemes \cite{10872781,10740049,9953071,jiang2024large}, our design enables flexible semantic decoupling. By leveraging cross-modal generative capabilities, the system can dynamically omit less important modalities when bandwidth is limited, relying on the remaining modality to generate the missing content.
At the receiver, reconstruction follows one of two adaptive workflows.
\begin{itemize}
	\item \textbf{Video-driven audio generation (V2A):} Designed for scenarios prioritizing visual fidelity, where text and 3DMM parameters are transmitted. The system first reconstructs the video content. The reconstructed video then guides the network to synthesize synchronized audio from the text, ensuring audiovisual synchronization.
	
	\item \textbf{Audio-driven video generation (A2V):} Designed for scenarios prioritizing audio accuracy, this system transmits only the audio semantics, including text and acoustic features such as duration, and first reconstructs the audio. The video generation network then synthesizes synchronized video based on the reconstructed audio.
\end{itemize}

\subsubsection{Technical Layer}
The technical layer manages the physical transmission of semantic features over the satellite channel. In addition, it provides real-time environmental information, such as satellite identification, user location, and weather conditions, to the LLM agent, enabling dynamic optimization of transmission strategies and network parameters.

\subsubsection{Semantic Knowledge Base}
The semantic knowledge base consists of explicit and implicit components. The explicit knowledge base stores relatively static information, such as user appearance, that is shared between the transmitter and receiver. Updates are required in scenarios involving user changes or significant pose variations. Due to satellite bandwidth constraints, these updates are strategically scheduled during periods of stable channel conditions. The implicit knowledge base is embedded within the semantic encoder and decoder through end-to-end training. It requires fine-tuning or retraining to adapt to evolving system scenarios.
The design and implementation details of each module are presented in the following sections.

\CheckRmv{\begin{figure*}[t]
		\centerline{\includegraphics[width=6.1in]{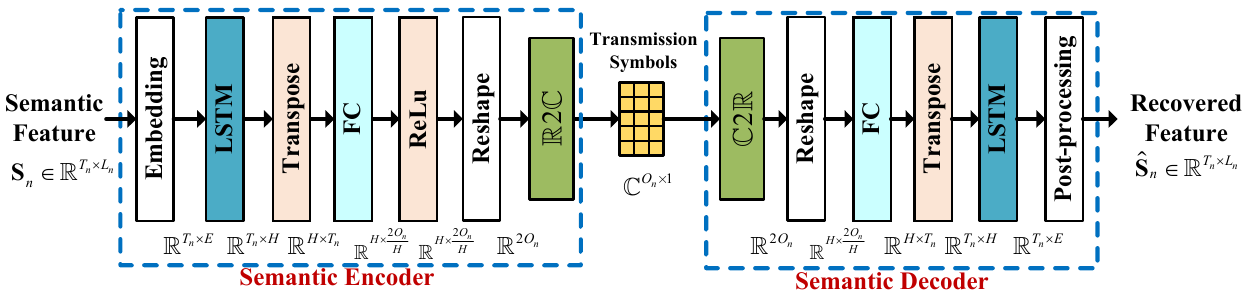}}
		\caption{Architecture of the temporal semantic encoder and decoder.} 
		\label{temporal codec}
\end{figure*}}

\subsection{Semantic Feature Extraction and Encoding/Decoding}
Facial dynamics in audiovisual content are mainly characterized by posture, expressions, and lip movements. Thus, a subset of 3DMM parameters \cite{booth20163d} is adopted as the key video semantics, significantly reducing data volume for bandwidth-constrained satellite links. At the receiver, these parameters, together with keyframes from the shared knowledge base, guide the generation of dynamic facial video.

Let $\mathbf{V} \in \mathbb{R}^{T_V \times 256 \times 256 \times 3}$ denote a video clip of length $T_V$, where the first frame $\mathbf{V}_1$ captures the user appearance. When the update condition is satisfied, this frame is used to update the knowledge base at both the transmitter and receiver. The subsequent frames are then processed by the 3DMM semantic extractor. The extraction process for the $i$th frame is given by
\CheckRmv{
	\begin{align} 
		\mathbf{s}_i = f_{\text{3DMM}}\!\left(\mathbf{V}_i; \Theta_{\text{3DMM}}\right),
		\label{eq10}
	\end{align}
}
where $f_{\text{3DMM}}(\cdot)$ is the pre-trained 3DMM extraction model \cite{deng2019accurate} with fixed parameters $\Theta_{\text{3DMM}}$. The vector $\mathbf{s}_i = [ \mathbf{s}_{i,\text{id}}, \mathbf{s}_{i,\text{exp}}, \mathbf{s}_{i,\text{rot}}, \mathbf{s}_{i,\text{trans}} ]$ comprises identity, expression, rotation, and translation coefficients. Since identity information is retained in the shared knowledge base, $\mathbf{s}_{i,\text{id}}$ is excluded from transmission. Furthermore, as the first six dimensions of $\mathbf{s}_{i,\text{exp}}$ capture most of the expression variations, only these coefficients are transmitted. Therefore, the transmitted semantic set for the $i$th frame is denoted as
\CheckRmv{\begin{equation}
		\mathbf{S}_{i,\text{M}}=\{\mathbf{s}_{i,\text{exp}}[1\!:\!6], \mathbf{s}_{i,\text{rot}}, \mathbf{s}_{i,\text{trans}}\},\label{eq11}
\end{equation}}
while the aggregated semantics for the entire video are denoted as $\mathbf{S}_{\text{M}}$.

For the audio signal $\mathbf{A}$ aligned with the video, we extract deep semantic features, including text, phonemes, and duration. Text conveys linguistic content, while phonemes and duration capture the acoustic characteristics of the audio. The extraction process is formulated as
\CheckRmv{\begin{equation}
		\mathbf{S}_{\text{T}}, \mathbf{S}_{\text{P}}, \mathbf{S}_{\text{D}} =
	f_{\text{audio}}(\mathbf{A}; \Theta_{\text{audio}}),
		\label{eq12}
\end{equation}}
where $\mathbf{S}_{\text{T}}$, $\mathbf{S}_{\text{P}}$, and $\mathbf{S}_{\text{D}}$ denote the extracted text, phonemes, and duration, respectively. The function $f_{\text{audio}}(\cdot)$ employs Whisper-small \cite{radford2023robust} for automatic speech recognition and the Montreal Forced Aligner \cite{mcauliffe2017montreal} for acoustic feature extraction.

Since $\mathbf{S}_{\text{M}}$, $\mathbf{S}_{\text{T}}$, $\mathbf{S}_{\text{P}}$, and $\mathbf{S}_{\text{D}}$ are time-dependent sequences, a temporal encoder-decoder architecture, as illustrated in \figref{temporal codec}, is adopted to mitigate the impact of satellite channel fading. The encoding process is given by
\CheckRmv{\begin{equation}
		\mathbf{X}_n = f_{\text{en},n}(\mathbf{S}_n; \Theta_{\text{en},n}),
	\quad n \in \{\text{M}, \text{T}, \text{P}, \text{D}\},
		\label{eq13}
\end{equation}}
where $\mathbf{X}_n$ is the encoded symbol, while $f_{\text{en},n}(\cdot)$ and $\Theta_{\text{en},n}$ represent the corresponding semantic feature encoder and its parameters, respectively.
At the receiver, features are reconstructed through the corresponding semantic decoders as
\CheckRmv{\begin{equation}
		\widehat{\mathbf{S}}_n = f_{\text{de},n}(\widehat{\mathbf{X}}_n; \Theta_{\text{de},n}),
	\quad n \in \{\text{M}, \text{T}, \text{P}, \text{D}\},
		\label{eq14}
\end{equation}}
where $\widehat{\mathbf{S}}_n$ denotes the recovered semantic features, and $f_{\text{de},n}(\cdot)$ is the decoder operation with parameters $\Theta_{\text{de},n}$.

We distinguish between floating-point data ($\mathbf{S}_{\text{M}}$, $\mathbf{S}_{\text{D}}$) and token sequences ($\mathbf{S}_{\text{T}}$, $\mathbf{S}_{\text{P}}$), and apply distinct encoding and decoding strategies, as illustrated in \figref{temporal codec}. For floating-point inputs, the embedding layer employs linear projection to map $\mathbb{R}^{T_n \times L_n}$ to $\mathbb{R}^{T_n \times E}$, where $T_n$ is the temporal length, $L_n$ is the input feature dimension, and $E$ is the embedding dimension. 
For token sequences, a lookup table $\boldsymbol{\Psi}_n$ encodes tokens into embeddings, which are further projected to $\mathbb{R}^{T_n \times E}$.
In the decoder, floating-point outputs are recovered using a single fully-connected layer, whereas token sequence outputs are obtained through a fully-connected layer followed by a softmax function and an argmax operation.

The training process for these encoder-decoder models is defined as
\CheckRmv{\begin{equation} 
		({{\widehat{\Theta }}_{\text{en},n}},{{\widehat{\Theta }}_{\text{de},n}})=\underset{{{\Theta }_{\text{en},n}},{{\Theta }_{\text{de},n}}}{\mathop{\arg \! \min }}\,{{\cal L}_{n}}({{\mathbf{S}}_{n}},{{f}_{\text{de,n}}}({{f}_{\text{en},n}}({{\mathbf{S}}_{n}};{{\Theta }_{\text{en},n}});{{\Theta }_{\text{de,n}}})),
		\label{eq15}
\end{equation}}
where ${\cal L}_{n}$ is the loss function. Specifically, mean-squared error (MSE) is employed for $\mathbf{S}_{\text{M}}$ and $\mathbf{S}_{\text{D}}$, while cross-entropy loss is utilized for $\mathbf{S}_{\text{T}}$ and $\mathbf{S}_{\text{P}}$.

\subsection{Multimodal Synchronous Generation Network}
Due to limited satellite bandwidth, the system uses a task-dependent selective transmission strategy. Instead of transmitting full multimodal data, only the essential semantic features required for the task are transmitted. At the receiver, these semantics are used to reconstruct the primary modality and guide the generation of the omitted modality. These two workflows are detailed below.

\CheckRmv{\begin{figure*}[!htp]
		\centerline{\includegraphics[width=6.8in]{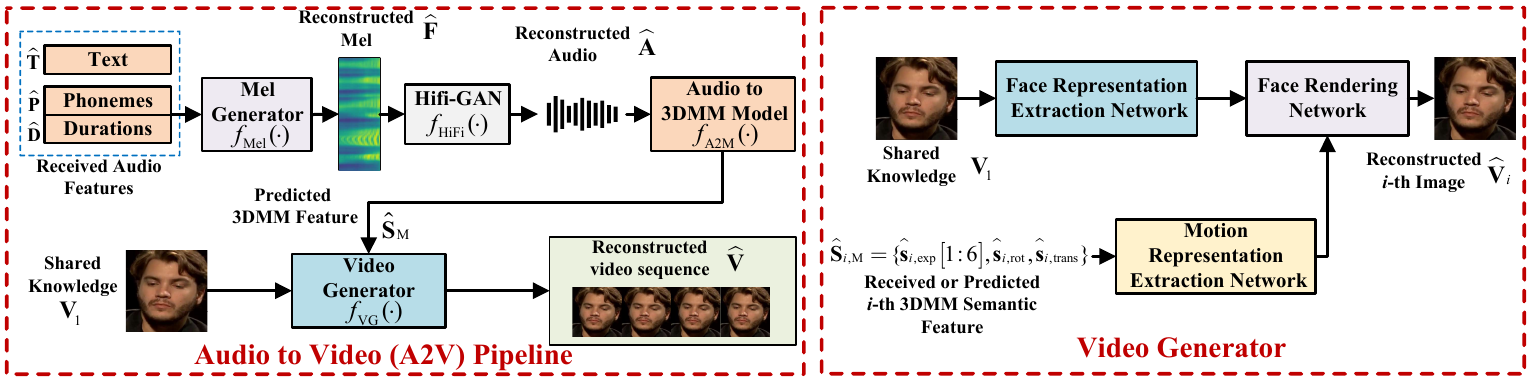}}
		\caption{Pipeline of the audio-driven video generation (A2V) and the structure of the video generator.} 
		\label{a2v}
\end{figure*}}

\CheckRmv{\begin{figure*}[!htp]
		\centerline{\includegraphics[width=6.8in]{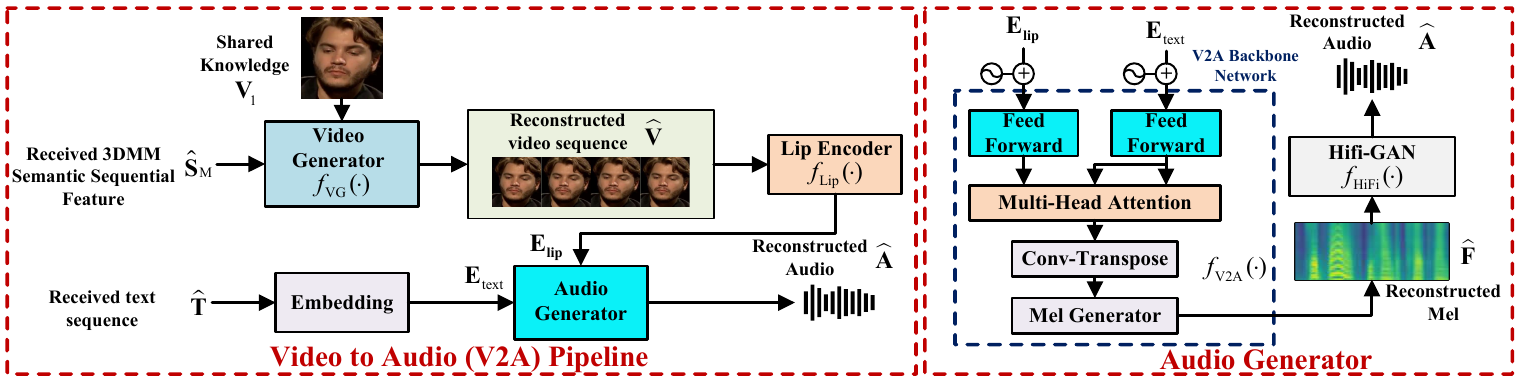}}
		\caption{Pipeline of the video-driven audio generation (V2A) and the structure of the audio generator.} 
		\label{v2a}
\end{figure*}}

\subsubsection{Audio-Driven Video Generation (A2V)} 
This workflow is designed for scenarios prioritizing audio accuracy, while audio semantics such as text, phonemes, and duration are transmitted. The received semantics are first used to reconstruct the audio $\widehat{\mathbf{A}}$, as illustrated in \figref{a2v}. The reconstructed phonemes $\widehat{\mathbf{S}}_{\text{P}}$ and duration $\widehat{\mathbf{S}}_{\text{D}}$ are processed by the Mel-spectrogram generator $f_{\text{Mel}}(\cdot)$ \cite{renfastspeech}, followed by the vocoder $f_{\text{HiFi}}(\cdot)$ \cite{kong2020hifi}, to reconstruct the audio waveform, which can be expressed as
\CheckRmv{\begin{equation} 
		\widehat{\mathbf{A}}={{f}_{\text{HiFi}}}\left( {{f}_{\text{Mel}}}\left( {{\widehat{\mathbf{S}}}_{\text{P}}},{{\widehat{\mathbf{S}}}_{\text{D}}};{{\Theta }_{\text{Mel}}} \right);{{\Theta }_{\text{HiFi}}} \right),
		\label{eq16}
\end{equation}}
where $\Theta_{\text{Mel}}$ and $\Theta_{\text{HiFi}}$ denote the parameters of $f_{\text{Mel}}(\cdot)$ and $f_{\text{HiFi}}(\cdot)$, respectively.

Subsequently, $\widehat{\mathbf{A}}$ is fed into the pre-trained audio-to-3DMM module $f_{\text{A2M}}(\cdot)$ \cite{ye2024real3d} to predict the 3DMM coefficients $\widehat{\mathbf{S}}_{\text{M}}$, which is formulated as
\CheckRmv{\begin{equation} 
		{{\widehat{\mathbf{S}}}_{\text{M}}}={{f}_{\text{A2M}}}\left( \widehat{\mathbf{A}};{{\Theta }_{\text{A2M}}} \right),
		\label{eq17}
\end{equation}}
where $\Theta_{\text{A2M}}$ denotes the parameters of $f_{\text{A2M}}(\cdot)$. This module first maps the audio signal into an audio feature representation and then predicts the corresponding 3DMM coefficients. As the module is trained on a large-scale audio-3DMM paired dataset, it effectively learns the mapping from audio to facial deformation.

Finally, the video generator $f_{\text{VG}}(\cdot)$ \cite{ye2024real3d} synthesizes consecutive video frames, as illustrated in \figref{a2v}. For the $i$th frame with coefficients $\widehat{\mathbf{S}}_{i,\text{M}}$, the generation process is defined as
\CheckRmv{\begin{equation} 
		{{\widehat{\mathbf{V}}}_{i}}={{f}_{\text{VG}}}\left( {{\widehat{\mathbf{S}}}_{i,M}},{{\mathbf{V}}_{1}};{{\Theta }_{\text{VG}}} \right),
		\label{eq18}
\end{equation}}
where $\mathbf{V}_{1}$ is the reference image that provides identity information from the knowledge base. Specifically, the motion representation extraction network converts the facial deformation and pose information in $\widehat{\mathbf{S}}_{i,\text{M}}$ into motion control features. These features, together with the static appearance features provided by $\mathbf{V}_{1}$, are then fed into the rendering network to animate the reference image, thereby generating the $i$th video frame $\widehat{\mathbf{V}}_{i}$.

\subsubsection{Video-Driven Audio Generation (V2A)} 
This workflow is designed for scenarios prioritizing visual fidelity, where text and 3DMM parameters are transmitted. As illustrated in \figref{v2a}, the received 3DMM parameters $\widehat{\mathbf{S}}_{\text{M}}$ and the shared reference image $\mathbf{V}_{1}$ are fed into the video generator $f_{\text{VG}}(\cdot)$ to synthesize the video sequence according to \eqref{eq18}.

Audiovisual synchronization is reflected in the coordination between lip movements and audio content. Therefore, the generated video ${{\widehat{\mathbf{V}}}}$ is used to guide the generation of time-aligned audio. Specifically, the lip region of the generated video is processed by a lip encoder $f_\text{Lip}(\cdot)$ \cite{martinez2020lipreading} to extract lip motion features, which can be expressed as
\CheckRmv{\begin{equation}
		{{\mathbf{E}}_{\text{lip}}}=f_\text{Lip}(\widehat{\mathbf{V}})\in {{\mathbb{R}}^{{{T}_{V}}\times {{D}_{m}}}},
		\label{eq19}	
\end{equation}}
where ${D}_{m}=256$ is the dimension of the lip features.

The received text is converted into embedding vectors ${{\mathbf{E}}_{\text{text}}}\in {{\mathbb{R}}^{1\times {{D}_{m}}}}$. A multi-head attention mechanism is then employed to learn the correlation between lip shapes and text embeddings \cite{cong2023learning}, guiding the audio generation process, which is formulated as
\CheckRmv{
	\begin{align}
		{{\mathbf{E}}_{\text{lip-text}}} &= \text{Attention}({{\mathbf{E}}_{\text{lip}}},{{\mathbf{E}}_{\text{text}}},{{\mathbf{E}}_{\text{text}}}) \notag \\
		&=\text{Softmax}\left( \frac{{{\mathbf{E}}_{\text{lip}}}{{\mathbf{E}}^{\top}_{\text{text}}}}{\sqrt{{{D}_{m}}}} \right){{\mathbf{E}}_{\text{text}}},
		\label{eq20}
	\end{align}	
}
where ${{\mathbf{E}}_{\text{lip-text}}} \in {{\mathbb{R}}^{{{T}_{V}}\times {{D}_{m}}}}$ is the aligned lip-text features, $\top$ is the transpose operation. According to \cite{cong2023learning}, in synchronized audiovisual clips, the Mel-spectrogram length $T_{\text{Mel}}$ is required to be $f_\text{expan}$ times that of the video frames $T_{V}$. Therefore, $\mathbf{E}_{\text{lip-text}}$ is expanded by a factor of $f_\text{expan}$ using transposed convolution. The expansion factor $f_\text{expan}$ is defined as
\CheckRmv{\begin{equation} 
		f_\text{expan}=\frac{{{T}_{Mel}}}{{{T}_{V}}}=\frac{sr/hs}{FPS}\in {{\mathbb{N}}^{+}},
		\label{eq21}	
\end{equation}} 
where $sr$, $hs$, and $FPS$ denote the audio sampling rate, the hop size used for Mel-spectrogram calculation, and the video frame rate, respectively.

The aligned features $\mathbf{E}_{\text{lip-text}}$ are then converted into a Mel-spectrogram sequence $\widehat{\mathbf{F}}$ by the generator $f_{\text{Mel}}(\cdot)$, and the audio waveform is synthesized by the vocoder $f_{\text{HiFi}}(\cdot)$. The synthesis process is expressed as
\CheckRmv{
	\begin{equation} 
		{\widehat{\mathbf{A}}}={{f}_{\text{HiFi}}}{\left( \widehat{\mathbf{F}};{{\Theta }_{\text{HiFi}}} \right)}={{f}_{\text{HiFi}}}\left( {{f}_{\text{Mel}}}\left( {{\mathbf{E}}_{\text{lip-text}}};{{\Theta }_{\text{Mel}}} \right);{{\Theta }_{\text{HiFi}}} \right).
		\label{eq22}
	\end{equation}
}

Let $f_{\text{V2A}}(\cdot)$ denote the V2A backbone network, which consists of the attention module, transposed convolution, and Mel-spectrogram generator, with parameters $\Theta_{\text{V2A}}$, as illustrated in \figref{v2a}. The network is trained by minimizing the discrepancy between the synthesized audio and the ground-truth audio, which can be expressed as
\CheckRmv{
	\begin{equation}
		{{\widehat{\Theta }}_{\text{V2A}}}=\underset{{{\Theta }_{\text{V2A}}}}{\mathop{\arg \! \min }} \left( \|\widehat{\mathbf{P}}-\mathbf{P}\|_2^2 + \|\widehat{\mathbf{E}}-\mathbf{E}\|_2^2 + \|\widehat{\mathbf{F}}-\mathbf{F}\|_F^2 \right),
		\label{eq23}
	\end{equation}
}
where $\mathbf{P}$, $\mathbf{E}$, and $\mathbf{F}$ denote the ground-truth audio's pitch, energy, and Mel-spectrogram, while $\widehat{\mathbf{P}}$, $\widehat{\mathbf{E}}$, and $\widehat{\mathbf{F}}$ are the predicted features. $\|\cdot\|_2$ and $\|\cdot\|_F$ represent the Euclidean norm and Frobenius norm, respectively.

\subsection{Semantic Knowledge Base Update Mechanism}
Current image-to-video generation semantic transmission frameworks typically rely on a single reference frame as a static shared knowledge base to provide appearance priors. While this ensures a certain degree of appearance consistency, it fails to adapt to variations such as pose changes, illumination shifts, or background transitions. As generation progresses, the initial reference frame often becomes inconsistent with the current content. Consequently, relying on an outdated knowledge base leads to significant distortion and quality degradation. However, in satellite scenarios, frequent updates involving high dimensional image data impose a heavy burden on limited bandwidth and may introduce significant latency.

\CheckRmv{\begin{figure} [t]
		\centerline{\includegraphics[width=3.4in]{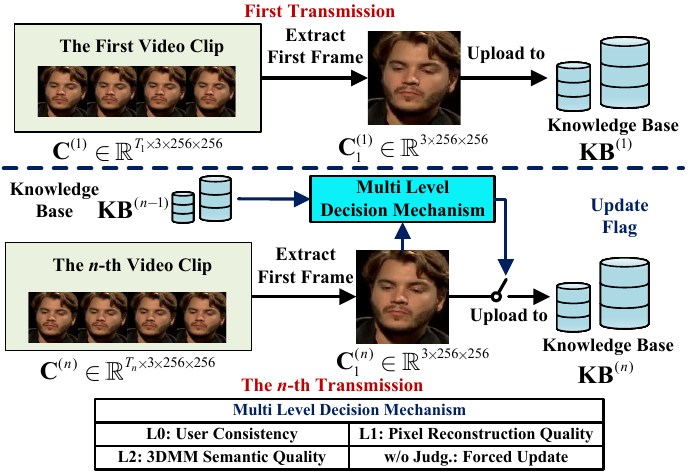}}
		\caption{Process of knowledge base update mechanism.}
		\label{kbm}
\end{figure}}

To address these issues, we propose an adjustable shared knowledge base construction and update mechanism that transmits new reference frames only when significant differences are detected, thereby optimizing satellite resource usage, as illustrated in \figref{kbm}. Let $\mathbf{C}=[\mathbf{C}^{(1)},\mathbf{C}^{(2)},\ldots,\mathbf{C}^{(N)}]$ represent a user multi-segment video collection, where the $n$th segment $\mathbf{C}^{(n)}$ contains $T_n$ frames. Initially, the first frame $\mathbf{C}_{1}^{(1)}$ of the first segment $\mathbf{C}^{(1)}$ is designated as the reference frame and bound to a user ID to initialize $\mathbf{KB}^{(1)}$ at both the transmitter and receiver. During decoding, the receiver retrieves this image via the user ID to guide the video synthesis process.

For the $n$th segment $\mathbf{C}^{(n)}$, the system evaluates differences between its first frame $\mathbf{C}_{1}^{(n)}$ and the reference images stored in the current $\mathbf{KB}^{(n-1)}$ to determine whether an update is required. To ensure efficiency and robustness, a multi-level decision mechanism is designed to evaluate identity consistency, low-level visual quality, and 3DMM semantic consistency.

\subsubsection{\textnormal{\textbf{L0: User Consistency Level}}} 
To maintain identity stability, this level evaluates reference frames in a face embedding space across different video segments. Specifically, the cosine similarity of identity embeddings (CSIM) \cite{deng2019arcface} is calculated to assess the similarity between the initial frame $\mathbf{C}_{1}^{(n)}$ of the current segment and the $i$th reference image $\mathbf{I}_i$ in the knowledge base, given by
\CheckRmv{
	\begin{equation}
		\text{CSIM}_i =
	\frac{F(\mathbf{C}_{1}^{(n)}) \cdot F(\mathbf{I}_i)}
	{\|F(\mathbf{C}_{1}^{(n)})\| \cdot \|F(\mathbf{I}_i)\|},
	\quad i \in \{1,2,\ldots,k\},
		\label{eq24}
	\end{equation}
}
where $F(\cdot)$ denotes the face embedding model, $\|\cdot\|$ represents the vector norm, and $k$ is the total number of images stored in the knowledge base.

Let $\text{CSIM}_i$ denote the maximum similarity between the current frame and all reference images in the knowledge base. If $\text{CSIM}_i$ exceeds the threshold $\alpha_{\text{CSIM}}$, the system considers the current frame sufficiently similar to $\mathbf{I}_i$ and reuses $\mathbf{I}_i$ for video generation. Otherwise, $\mathbf{C}_{1}^{(n)}$ is appended to the knowledge base as a new reference frame.

\CheckRmv{\begin{figure*}[!ht]
		\centerline{\includegraphics[width=6.3in]{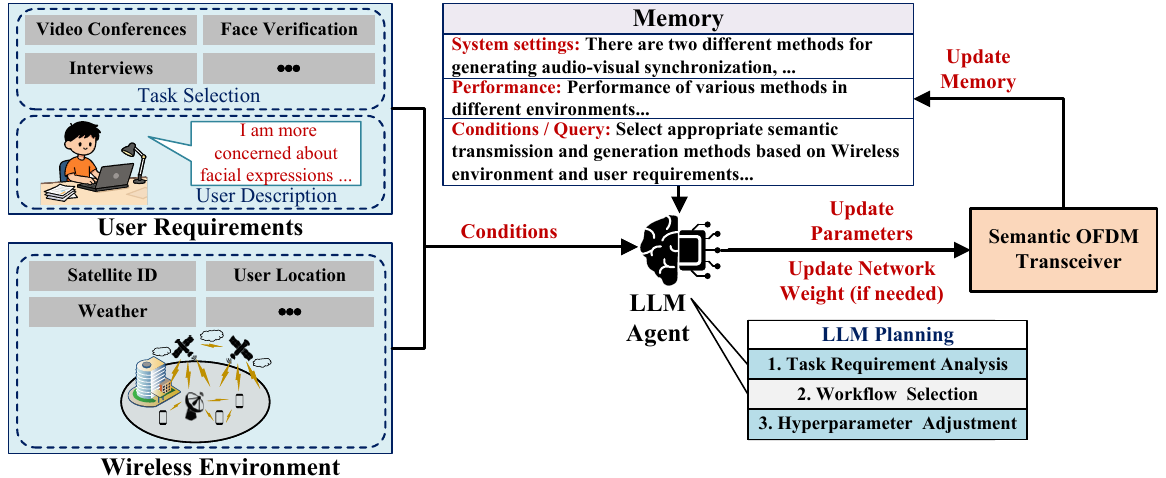}}
		\caption{Adaptive task- and environment-aware strategy based on an LLM agent.} 
		\label{agent}
\end{figure*}}

\subsubsection{\textnormal{\textbf{L1: Pixel Reconstruction Quality Level}}}
This level evaluates low-level visual consistency, such as brightness and texture, using the peak signal-to-noise ratio (PSNR), which is computed as
\CheckRmv{
\begin{align}
	\text{PSNR}_i
	&= 10 \log_{10}
	\left(
	\frac{255^2}{\text{MSE}(\mathbf{C}_{1}^{(n)}, \mathbf{I}_i)}
	\right),
	\quad i \in \{1,2,\ldots,k\},
	\label{eq25}
\end{align}}
where $\text{MSE}(\cdot)$ denotes the MSE.

Let $\mathbf{I}_i$ be the reference image that maximizes PSNR with respect to $\mathbf{C}_{1}^{(n)}$. Provided that the L0 condition is satisfied, the system compares $\text{PSNR}_i$ with the threshold $\alpha_{\text{PSNR}}$ in dB. If $\text{PSNR}_i > \alpha_{\text{PSNR}}$, $\mathbf{I}_i$ is reused as the reference frame. Otherwise, $\mathbf{C}_{1}^{(n)}$ is added to the knowledge base.

\subsubsection{\textnormal{\textbf{L2: 3DMM Semantic Quality Level}}} 
This level evaluates three-dimensional geometric discrepancies, including pose, expression, and translation. Using 3DMM fitting, the system extracts 3D parameters and quantifies the disparity between the current frame and reference images. The weighted 3DMM parameter distance between $\mathbf{I}_i$ and $\mathbf{C}_{1}^{(n)}$ is defined as
\CheckRmv{
	\begin{align}
		{{d}_{i}}={{\lambda }_{\exp }}{{d}_{i,\text{exp}}}+{{\lambda }_{\text{rot}}}{{d}_{i,\text{rot}}}+{{\lambda }_{\text{trans}}}{{d}_{i,\text{trans}}},
		\label{eq26}
	\end{align}	
}
where
$d_{i,\text{exp}}=\|\boldsymbol{\delta}_{\text{exp}}^{\mathbf{C}_{1}^{(n)}}-\boldsymbol{\delta}_{\text{exp}}^{\mathbf{I}_i}\|$,
$d_{i,\text{rot}}=\|\boldsymbol{\delta}_{\text{rot}}^{\mathbf{C}_{1}^{(n)}}-\boldsymbol{\delta}_{\text{rot}}^{\mathbf{I}_i}\|$, and
$d_{i,\text{trans}}=\|\boldsymbol{\delta}_{\text{trans}}^{\mathbf{C}_{1}^{(n)}}-\boldsymbol{\delta}_{\text{trans}}^{\mathbf{I}_i}\|$,
with weighting factors $\lambda_{\text{exp}}$, $\lambda_{\text{rot}}$, and $\lambda_{\text{trans}}$.

Let $d_i$ be the minimum distance among all reference images. Provided that both L0 and L1 conditions are satisfied, the system verifies whether
$d_{i,\text{exp}} < \alpha_{\text{exp}}$,
$d_{i,\text{rot}} < \alpha_{\text{rot}}$, and
$d_{i,\text{trans}} < \alpha_{\text{trans}}$.
If all conditions are met, $\mathbf{I}_i$ is reused. Otherwise, $\mathbf{C}_{1}^{(n)}$ is added to the knowledge base.

\subsubsection{\textnormal{\textbf{w/o Judg.: Forced Update Level}}}
This level assumes sufficient available bandwidth. The system bypasses all similarity evaluations and directly adds $\mathbf{C}_{1}^{(n)}$ to the knowledge base as a new reference frame for each segment. Although this strategy maximizes reconstruction quality and temporal consistency, it incurs significantly higher bandwidth overhead. 

\subsection{LLM Agent Based Environment and Task Adaptability}
Although the proposed dual-stream generation framework and knowledge base update mechanism provide adaptability to user and scenario requirements, real-time configuration adjustment remains challenging due to fluctuating channel conditions and varying task types. In complex environments characterized by high-mobility dynamics and rain attenuation, static configurations often fail to ensure optimal performance. To address this limitation, we propose a system capable of automatically generating optimized configuration schemes and workflows tailored to channel environments, satellite-to-ground bandwidth constraints, and user requirements.

Conventional rule-based or lookup-table methods are effective in simple scenarios but suffer from combinatorial explosion as task diversity increases, making maintenance and scalability difficult. Moreover, these methods often fail to capture the requirements underlying different tasks. In contrast, LLMs possess inherent semantic understanding and autonomous planning capabilities, enabling effective interaction with both users and the environment. Therefore, an LLM-based agent is introduced for adaptive semantic transmission decision making in satellite communication systems. This agent enables the dynamic selection of semantic strategies and bandwidth configurations based on multidimensional environmental information and user requirements.

During the initialization phase, the LLM agent is configured through prompt engineering. It analyzes system configurations and historical transmission logs, such as satellite identifiers, orbital positions, and weather conditions, as well as the performance of different semantic strategies. By learning the mapping between wireless environments and task performance, the agent constructs an initial memory base to support subsequent decision making.

During operation, the agent processes inputs including task descriptions, user preferences, such as quality priority, and real-time environmental data. It then executes a three-step reasoning and planning procedure.
\begin{itemize}
	\item \textbf{Task requirement analysis:} The agent evaluates the current satellite link quality using environmental information and analyzes the service requirements under the current link condition according to the explicitly specified transmission task and user preferences.
	
	\item \textbf{Workflow selection:} Based on the inferred context, the agent selects the most suitable workflow from the supported generation workflows and semantic transmission strategies.
	
	\item \textbf{Resource and hyperparameter adjustment:} The agent dynamically adjusts hyperparameters, including the semantic feature compression rate, bandwidth allocation, and the activation of knowledge base updates, to ensure task completion.
\end{itemize}

The resulting plan directly configures the satellite OFDM transceiver. After task completion, the system records the full transmission context in the memory base. If the LLM detects configuration inadequacy during inference, it generates new candidate schemes. Once verified, these schemes are incorporated into the configuration set, enabling the system to continuously evolve and adapt to diverse service scenarios. 

\begin{table}[t]
	\centering
	\caption{System Parameters}
	\label{tab:link_params}
	\begin{tabular}{lc}
		\toprule
		Parameter & Value \\
		\midrule
		Transmit Power ($P_\text{t}$) & 25 dBm \\
		Transmit Antenna Gain ($G_\text{t}$) & 40 dBi \\
		Satellite Equivalent Gain ($G_\text{s}$) & 85 dBi \\
		Receive Antenna Gain ($G_\text{r}$) & 40 dBi \\
		Boltzmann Constant ($k$) & 1.38e-23 J/K \\
		Noise Temperature ($\mathcal{T}$) & 290 K \\
		System Bandwidth ($B$) & 20 MHz \\
		\bottomrule
	\end{tabular}
\end{table}

\section{Numerical Results}\label{section:Numerical Results}
In this section, we present a performance and bandwidth consumption comparison between the proposed system and existing audio and video transmission methods. Experiments were conducted utilizing the LRS2 \cite{8585066} and VoxCeleb \cite{nagrani2017voxceleb} datasets, which comprise time-aligned talking-head videos. The videos were resized to a resolution of 256$\times$256. The training set consists of 40,000 audiovisual pairs from LRS2, while the test set includes an additional 8,000 audiovisual pairs from LRS2 and a subset of the VoxCeleb dataset.

To simulate the satellite environment, we employed the NTN-TDL-A channel model. The satellite altitude was randomly selected from the range of 300 km to 1,200 km, and the ground terminal's movement speed was set to 3 km/h. The physical layer OFDM system comprised 14 symbols per frame and 120 subcarriers. A comb-type pilot pattern was implemented with a spacing of 4 subcarriers. Therefore, 30 subcarriers were allocated for pilots and 90 for data transmission. Other system parameters are shown in \tabref{tab:link_params}.

\subsection{Network Setting and Benchmarks}\label{Network Setting}
The deployment and training strategies for the framework's modules are detailed as follows:
\begin{itemize}
	\item \textbf{Semantic extraction module:} Comprising 3DMM extraction, speech recognition, and acoustic information extraction networks, these pre-trained models are deployed directly at the transmitter.
	
	\item \textbf{Semantic encoder and decoder:} Distinct encoder-decoder networks are established for each semantic feature type (3DMM, text, phonemes, duration). These networks are trained under noisy satellite channels to minimize transmission error, following \eqref{eq15}.
	
	\item \textbf{Multimodal reconstruction module:} The modules $f_\text{VG}(\cdot)$, $f_\text{A2M}(\cdot)$, $f_\text{Mel}(\cdot)$, $f_\text{HiFi}(\cdot)$, and $f_\text{Lip}(\cdot)$ utilize pre-trained weights from large-scale audiovisual datasets, which demonstrate robust performance. Conversely, the lip-to-Mel-spectrogram generator $f_\text{V2A}(\cdot)$ requires specific training via \eqref{eq23} to mitigate channel noise and ensure audiovisual synchronization, conditioned on the generated video $\widehat{\mathbf{V}}$ and received text.
	
	\item \textbf{LLM agent:} The system utilizes GPT-4o \cite{achiam2023gpt} as the decision-making agent, and no fine-tuning is performed. The agent generates decisions based on prompts and historical transmission-performance references.	
	
\end{itemize}

All networks requiring training utilized the Adam optimizer. Specifically, the four semantic encoder-decoder pairs were trained for 400 epochs with a learning rate of 0.001, while the $f_\text{V2A}(\cdot)$ was trained for 1000 epochs with a learning rate of 0.0001. The parameters for semantic knowledge base update mechanism are set to $\alpha_{\text{CSIM}}=0.7$, $\alpha_{\text{PSNR}}=13$, $\alpha_{\text{exp}}=8$, $\alpha_{\text{rot}}=0.3$, $\alpha_{\text{trans}}=0.3$, $\lambda_{\text{exp}}=0.1$, $\lambda_{\text{rot}}=1$, and $\lambda_{\text{trans}}=1$.

To compare performance, for video modality, we used H264/H265 with LDPC coding and 64 quadrature amplitude modulation (64-QAM) as traditional baselines. In addition, we included two representative semantic video baselines: SVC \cite{9955991}, which transmits facial keypoints for video generation at the receiver, and DeepWiVe \cite{9837870}, a representative JSCC-based video semantic transmission scheme that exploits temporal correlation across frames. For the audio modality, the end-to-end waveform encoding and decoding method DeepSC-S \cite{9450827} was used as the baseline for semantic audio transmission, and the text-to-audio generation network FastSpeech 2 \cite{renfastspeech} was employed as the baseline for audiovisual synchronization.

\begin{table}[!t]
	\caption{Computational Complexity and Transmission Bandwidth}
	\centering
	\scriptsize 
	\begin{tabular}{llrrr} 
		\toprule
		\textbf{Type} & \textbf{Method} & \textbf{\tabincell{r}{Transmission \\ Symbols}} & {\textbf{Params [M]}} & {\textbf{Runtime [s]}}  \\
		\midrule
		\multirow{5}{*}{Video Level}
		& H264+LDPC         & 400,991   & {--} & 0.33e-1 \\
		& H265+LDPC         & 54,390    & {--} & 0.13e-1 \\
		& DeepWiVe          & 59,392    & 0.89 & 0.97e-2 \\
		& SVC               & 600       & 60.11 & 0.19e-1 \\
		& V2A (Video Part)  & 300       & 172.01 & 0.71e-1 \\
		& A2V (Video Part)  & 0         & 159.88 & 0.53e-1 \\
		\midrule
		\multirow{4}{*}{Audio Level}
		& FastSpeech 2      & 600       & 70.46 & 0.19e-1 \\
		& DeepSC-S          & 32,768    & 0.75  & 0.21e-2 \\
		& V2A (Audio Part)  & 300       & 368.89 & 0.10e-0 \\
		& A2V (Audio Part)  & 600       & 317.25 & 0.65e-1 \\
		\bottomrule
	\end{tabular}
	\label{table:complexity}
\end{table}

As shown in \tabref{table:complexity}, conventional H.265 and DeepSC-S have low inference overhead due to their lightweight decoding pipelines, while DeepWiVe also maintains moderate computational complexity but requires substantially more transmission symbols. In contrast, the proposed V2A and A2V schemes incur higher latency and peak memory usage because of semantic extraction and cross-modal generation. The LLM module is invoked only at session initialization or under major environmental changes, so its overhead is amortized rather than accumulated per frame. Despite the increased computation, V2A and A2V achieve significant bandwidth savings over H.265, DeepSC-S, and DeepWiVe, with A2V even enabling zero-symbol video transmission by generating video from audio semantics. Therefore, the proposed framework offers a favorable computation-bandwidth trade-off for narrowband satellite links, where bandwidth is scarce and retransmissions are costly. In practical deployment, lightweight semantic encoding can run on mobile terminals, while generative reconstruction can be offloaded to ground or edge receivers, model pruning and quantization can further reduce the resource cost.

The performance of the multimodal semantic transmission is evaluated across three dimensions: video reconstruction, audio reconstruction, and audiovisual synchronization.

\begin{itemize}
	\item \textbf{Video reconstruction:} Video quality is assessed using the CSIM \cite{deng2019arcface} for identity consistency and the learned perceptual image patch similarity (LPIPS) \cite{Zhang_2018_CVPR} for perceptual discrepancy based on deep features. Given the focus on facial video, average keypoint distance (AKD) \cite{siarohin2019first} is employed to evaluate the accuracy of facial motion and expression. Here, higher CSIM and lower LPIPS or AKD indicate better quality.
	
	\item \textbf{Audio reconstruction:} 
	Evaluation considers both intelligibility and perceptual quality. Word error rate (WER) measures the accuracy of textual semantics. Perceptual quality is assessed via perceptual evaluation of speech quality (PESQ) \cite{rix2001perceptual}, an ITU-T P.862 standard evaluating audio clarity and naturalness. Lower WER and higher PESQ are preferred.
	
	\item \textbf{Audiovisual synchronization:} 
	We employ lip-sync error distance (LSE-D) and lip-sync error confidence (LSE-C) \cite{prajwal2020lip} to quantify lip-synchronization performance. LSE-D measures the temporal mismatch between lip movements and audio, while LSE-C evaluates synchronization reliability through confidence scores. Lower LSE-D and higher LSE-C denote better synchronization.
\end{itemize} 

\CheckRmv{\begin{figure*}[!t]
		\setkeys{Gin}{width=3.0in}  
		\centering
		\subfloat[]{\includegraphics{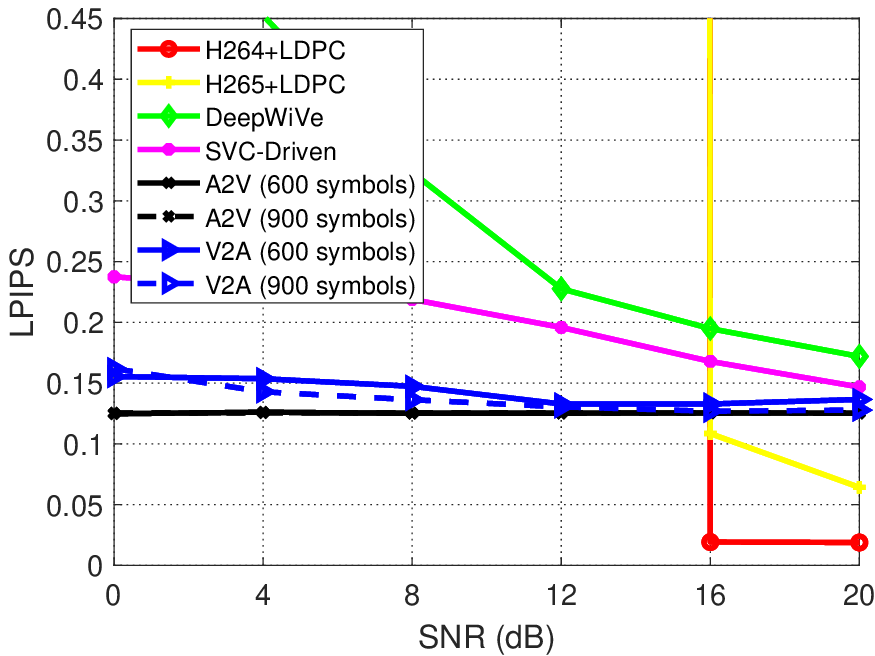}%
			\label{lpips}}
		\hfil
		\subfloat[]{\includegraphics{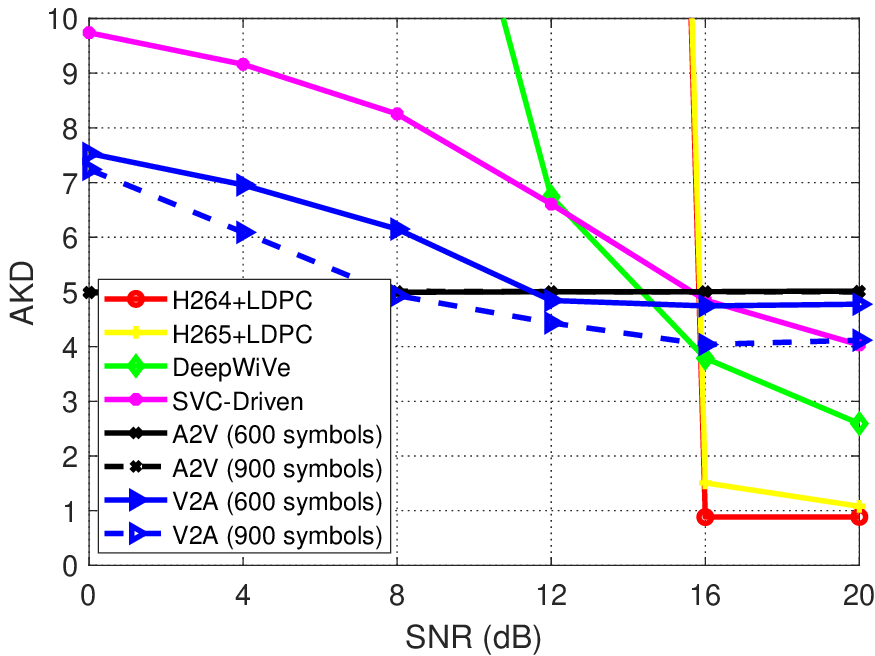}%
			\label{akd}	}
		\caption{Performance of different video transmission methods under varying SNR conditions. (a) LPIPS performance; (b) AKD performance.}
		\label{video performance}
\end{figure*}}

\subsection{Semantic Transmission Performance}\label{channel environments} 
This section evaluates the system performance in terms of video fidelity, audio quality, and audiovisual synchronization. \figref{video performance} compares video transmission methods using LPIPS and AKD. As shown in \figref{video performance}\subref{lpips}, under high SNR, traditional H264/H265 achieve lower LPIPS but require significantly higher bandwidth. However, under low SNR, traditional methods suffer from rapid performance degradation, whereas generative methods (A2V and V2A) exhibit superior robustness. While DeepWiVe partially mitigates the sharp performance degradation of traditional methods under low SNR conditions, its required transmission bandwidth remains comparable to that of H265 and significantly higher than that required by V2A and A2V. The AKD results in \figref{video performance}\subref{akd} confirm this trend. While traditional methods excel only at high SNR, V2A maintains stable keypoint reconstruction even at low SNR, underscoring its capability in preserving video semantics. 

To evaluate the impact of bandwidth on performance, we set the total bandwidth for A2V and V2A to 600 and 900 symbols. In both cases, 300 symbols are allocated to text, while the remaining bandwidth is used for video or audio features. With increased bandwidth, V2A (900 symbols) significantly improves AKD compared to V2A (600 symbols), demonstrating that higher bandwidth enhances video semantic reconstruction. In contrast, A2V, which only transmits audio semantics and drives video generation through audio, performs well under bandwidth constraints or poor channel conditions. However, it exhibits a performance bottleneck, where additional bandwidth does not further enhance video reconstruction. 

\begin{figure}[t]
	\centering
	\setkeys{Gin}{width=0.125\textwidth}  
	\subfloat[Original]{\includegraphics{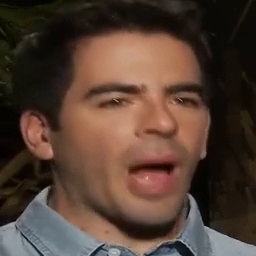}\label{picture1}} 
	\hspace{2mm}
	\subfloat[V2A]{\includegraphics{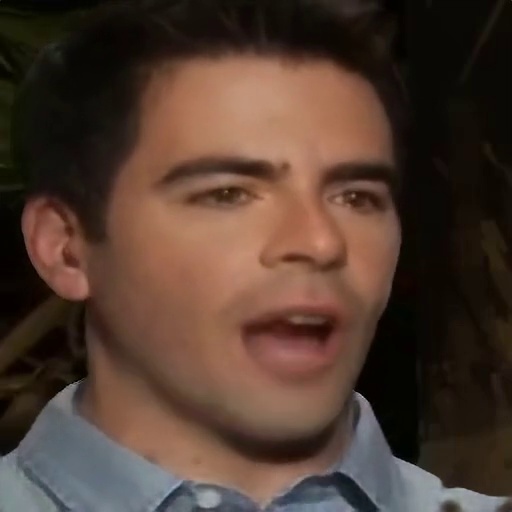}\label{picture2}}
	\hspace{2mm}
	\subfloat[A2V ]{\includegraphics{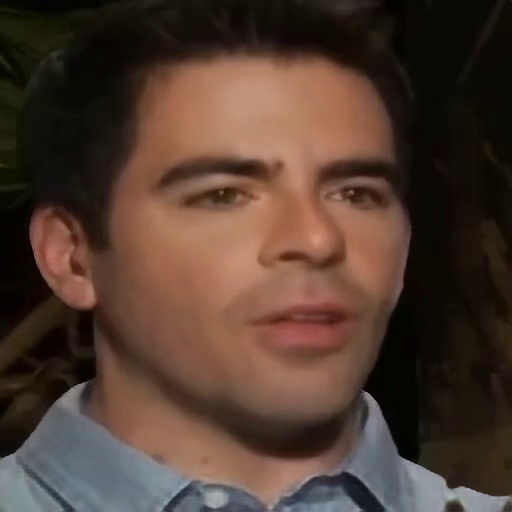}\label{picture3}} \\
	\subfloat[SVC ]{\includegraphics{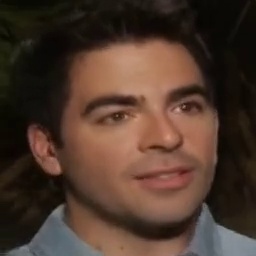}\label{picture4}}
	\hspace{2mm}
	\subfloat[H264 ]{\includegraphics{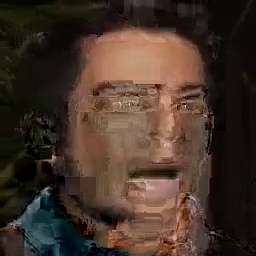}\label{picture5}} 
	\hspace{2mm}
	\subfloat[H265 ]{\includegraphics{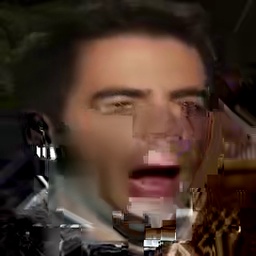}\label{picture6}}
	\caption{Reconstruction results of various methods under SNR = 12 dB. (a) Original image; (b) V2A (AKD = 5.41); (c) A2V (AKD = 5.85); (d) SVC (AKD = 8.36); (e) H264 (AKD = {N/A}); (f) H265 (AKD = {N/A}). N/A indicates that the keypoints cannot be detected to calculate AKD because the face is blurred.}
	\label{picture}
\end{figure}

\figref{picture} shows reconstructed video frames and AKD performance at 12 dB SNR. V2A achieves superior reconstruction by transmitting video semantics. In contrast, A2V is limited by its audio-driven video generation mechanism. SVC relies on geometric keypoints, which are extremely sensitive to channel noise. Even minor interference can lead to significant facial misalignment. In contrast, traditional methods fail to reconstruct recognizable content under the dual constraints of limited bandwidth and severe noise.

\tabref{table:speech} evaluates audio transmission methods at 20 dB SNR. The LSE-C and LSE-D are based on a comparison between the transmitted audio and the original video, with the original audio serving as the ground truth. DeepSC-S maintains high quality but consumes approximately $28\times$ the bandwidth of generative methods. A2V, by transmitting additional semantics such as duration, achieves synchronization scores comparable to DeepSC-S with minimal overhead. Meanwhile, FastSpeech 2 and V2A show a marginal decline in synchronization due to the absence of acoustic characteristics. It is also observed that FastSpeech 2 achieves the lowest WER by allocating the highest bandwidth to text.

\CheckRmv{\begin{figure}[t]
		\centering
		\setkeys{Gin}{width=2.8in}  
		\subfloat[]{\includegraphics{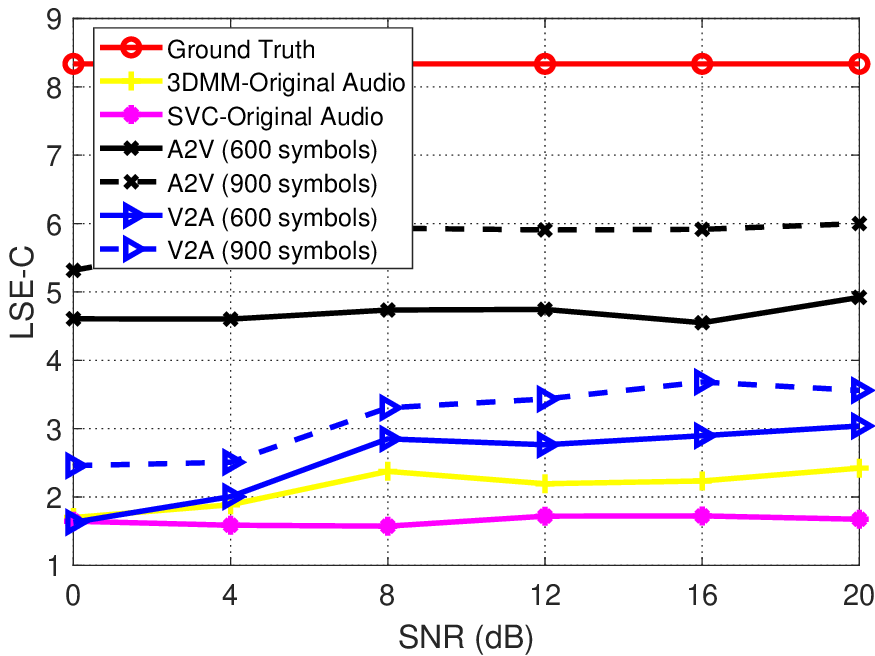}%
			\label{LSEC}}
		\hspace{10mm}
		\subfloat[]{\includegraphics{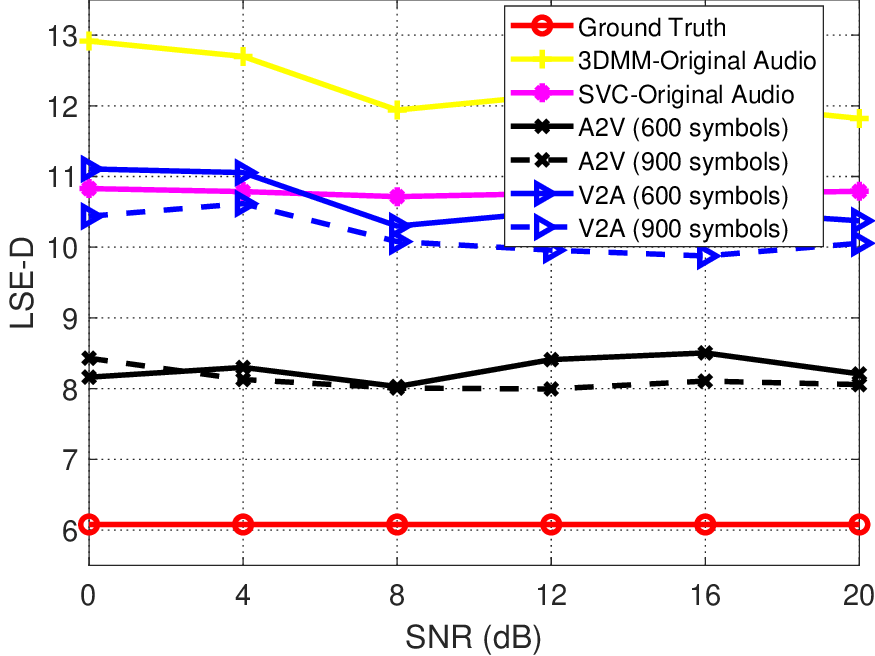}%
			\label{LSED}}
		\caption{Audiovisual synchronization performance of different transmission methods. (a) LSE-C performance; (b) LSE-D performance.} 
		\label{LSE}
\end{figure}}

\begin{table}[t]
	\caption{Audio Transmission Performance of Different Methods}
	\label{table:speech}
	\centering
	\begin{tabular}{lrrrr} 
		\hline
		\textbf{Method} & \textbf{LSE-C} & \textbf{LSE-D} & \textbf{WER} & \textbf{PESQ} \\
		\hline
		Original Audio  & 8.33 & 6.07 & 0.036 & 4.50 \\
		DeepSC-S  & 7.85 & 6.57 & 0.11 & 2.99 \\
		A2V  & 5.85 & 8.69 & 0.11 & 1.25 \\
		V2A  & 2.22 & 12.16 & 0.11 & 1.18 \\
		FastSpeech 2  & 3.12 & 10.21 & 0.079 & 1.15 \\
		\hline
	\end{tabular}
\end{table}

\figref{LSE}\subref{LSEC} and \subref{LSED} illustrate audiovisual synchronization performance. 3DMM and SVC, which generate video based on geometric parameters or keypoints, exhibit a low match with the original audio, resulting in poorer performance. In contrast, the cross-modal generative methods (A2V and V2A) demonstrate superior synchronization, with performance improving consistently as bandwidth increases. In summary, V2A prioritizes video semantic transmission, yielding significant quality gains with increased bandwidth, while A2V excels in audio-related semantic modeling. Therefore, in subsequent application stages, the proposed system can dynamically select the optimal transmission path based on channel conditions and task requirements, making it highly effective for narrowband satellite communication scenarios.

\CheckRmv{\begin{figure}[t]
		\centering
		\setkeys{Gin}{width=2.8in}  
		\subfloat[]{\includegraphics{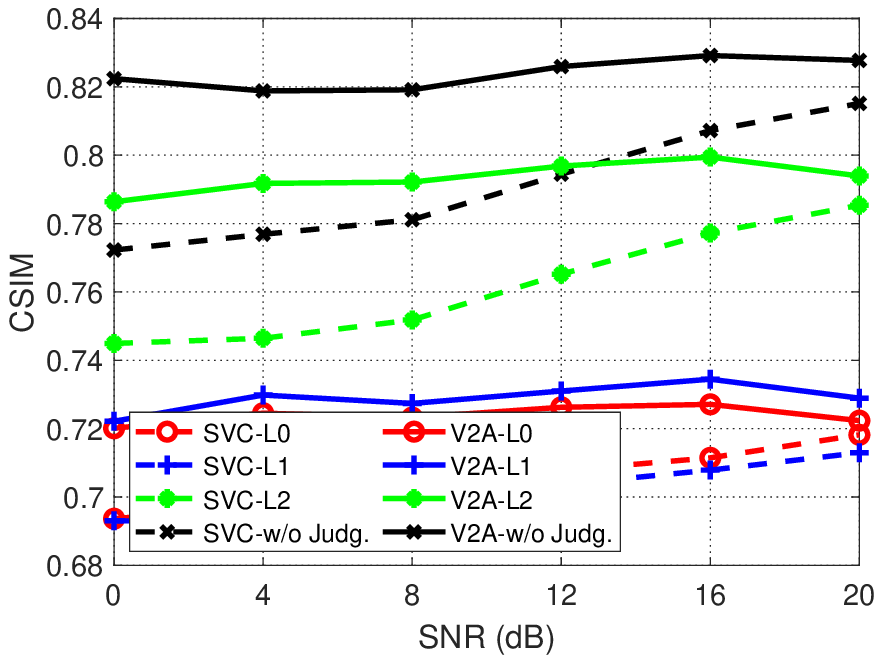}%
			\label{kbcsim}}
		\hspace{10mm}
		\subfloat[]{\includegraphics{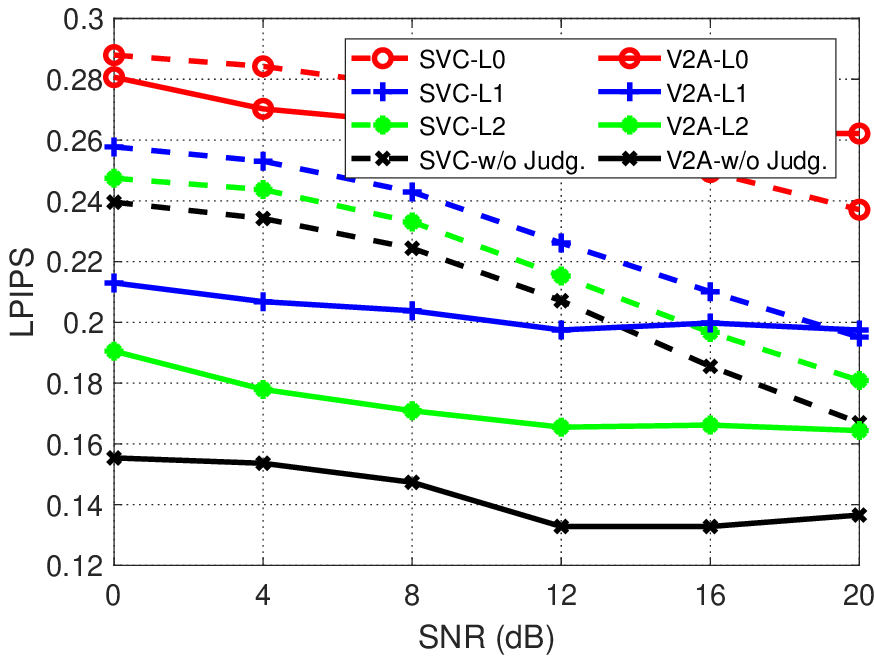}%
			\label{kblpisp}}
		\hspace{10mm}
		\subfloat[]{\includegraphics{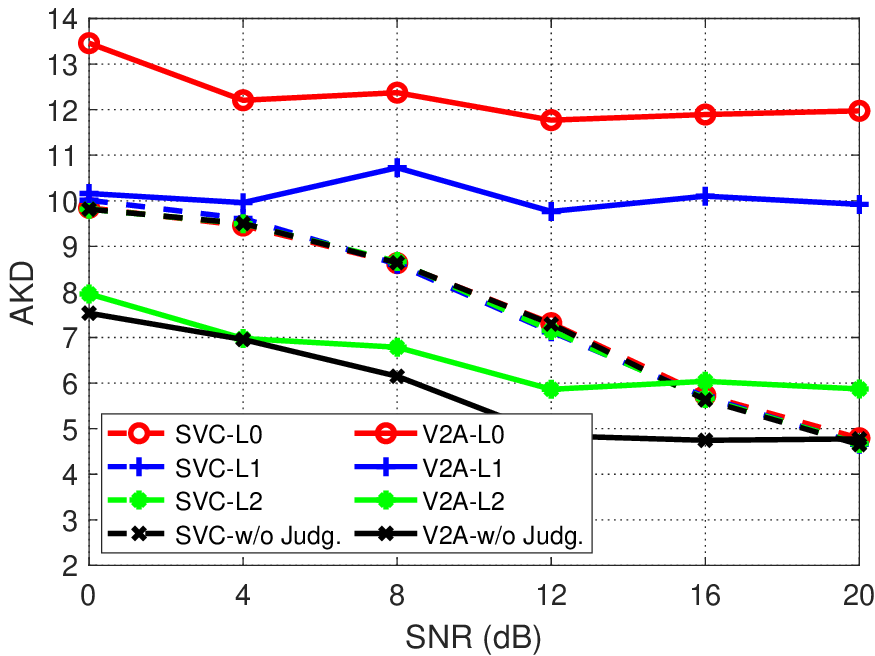}%
			\label{kbakd}}
		\caption{Impact of knowledge base update levels on different semantic transmission methods.
			(a) CSIM performance;
			(b) LPIPS performance;
			(c) AKD performance.} 
		\label{kb}
\end{figure}}

\subsection{Impact of Knowledge Base on Semantic Method Performance}
The quality of the shared knowledge base significantly influences generative performance. To assess the proposed update mechanisms, we conducted experiments in scenarios involving the transmission of 100 video segments with multiple users, measuring CSIM, LPIPS, and AKD, as shown in \figref{kb}. We compare the proposed L0-L2 against a baseline w/o Judg. (which updates the reference frame for every segment). \tabref{table:Kbb} shows the average bandwidth overhead for each level, where the semantic features (3DMM parameters or keypoints) have a fixed bandwidth, and user image updates in the knowledge base employ a JSCC image transmission method \cite{8723589} with a compression rate of 1/12, transmitted over a high-quality channel. The transmission of a 256$\times$256$\times$3 image requires 16,384 symbols.

\begin{table}[t]
	\centering
	\caption{Average Transmission Bandwidth Per Video Across Different Knowledge Base Update Levels} 
	\label{table:Kbb}
	\begin{tabular}{crrr} 
		\toprule
		\textbf{Update Level} & \textbf{Update Count} & \textbf{\tabincell{r}{Semantic \\ Symbols}} & \textbf{\tabincell{r}{Knowledge Base \\ Update Symbols}}\\
		\midrule
		L0 & 17 & 300 & 2,785 \\
		L1 & 27 & 300 & 4,427 \\
		L2 & 50 & 300 & 8,192 \\
		w/o Judg. & 100 & 300 & 16,384 \\
		\bottomrule
	\end{tabular}
\end{table}

These results demonstrate that performance gradually improves as the judgment criteria become more stringent. Specifically, L0 (identity-only) maintains CSIM $>0.7$, ensuring basic identity recognition, but performs poorly in texture details and geometric accuracy, resulting in mediocre LPIPS and AKD scores. Integrating pixel-level (L1) and semantic-level (L2) constraints significantly enhances reconstruction accuracy for both SVC and V2A. In contrast, w/o Judg. shows slight advantages across all metrics, but its update frequency and bandwidth consumption increase dramatically. In satellite links characterized by high latency and error rates, such frequent high-volume transmissions pose a risk of link congestion and reliability issues, leading to deployment bottlenecks. Notably, V2A-L2 closely approximates the performance of V2A-w/o Judg.. For example, at 12 dB SNR, V2A-L2 achieves an AKD of 5.8, comparable to the 4.8 of w/o Judg., while consuming only about 50\% of the bandwidth required by w/o Judg.. This demonstrates that the proposed mechanism effectively balances generation quality with resource conservation, ensuring stable and cost-effective multimodal transmission in bandwidth-limited satellite environments.

\CheckRmv{\begin{figure} [t]
		\centerline{\includegraphics[width=2.9in]{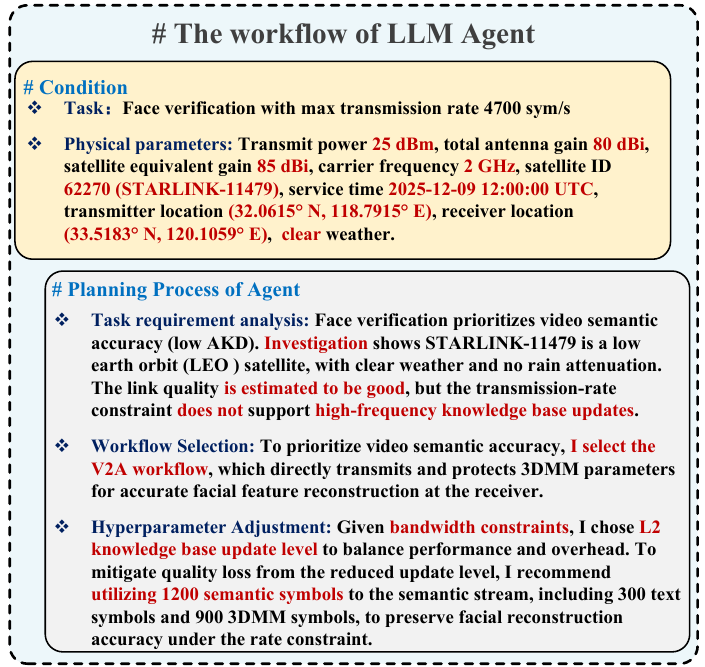}}
		\caption{Decision Process of the LLM agent for the face-verification task under a transmission-rate constraint of 4700 sym/s.}
		\label{decision}
\end{figure}}

\subsection{Case Study on Task and Environment Adaptability}
This case study evaluates the practical deployment of the LLM agent, utilizing the simulation results from previous sections as the agent's long-term memory. The agent's performance is compared with the traditional ``Lookup-Table'' method, which operates via static, rule-based queries of historical performance data. The agent's reasoning process for a face-verification task is detailed in \figref{decision}. The simulation scenario is defined by a transmit power of 25 dBm, total antenna gains of 80 dBi, a satellite equivalent gain of 85 dBi, and a 2 GHz carrier frequency, serviced by STARLINK-11479 (ID 62270). The evaluation period is centered around 2025-12-09 12:00:00 (UTC) for the transmitter at ($32.0615^\circ$ N, $118.7915^\circ$ E) and the receiver at ($33.5183^\circ$ N, $120.1059^\circ$ E) during clear weather.

Upon receiving these inputs, the agent first analyzes the task and scenario, recognizing the high video semantic accuracy requirement for face verification. By evaluating the physical parameters, the agent assesses the link quality as good but identifies ``bandwidth limitation'' as a core constraint, which restricts frequent knowledge base updates. Therefore, the agent selects the V2A workflow to prioritize direct transmission and protection of 3DMM parameters, ensuring accurate facial reconstruction. To optimize resource usage, it downshifts from the high-cost w/o Judg. update scheme to the more balanced L2. Additionally, the saved bandwidth is reallocated to preserve the accuracy of 3DMM semantics, thereby mitigating potential quality loss and ensuring service quality within resource limits.

\figref{llm} shows the AKD performance of different strategies over several 20-second intervals near the service time (2025-12-09 12:00:00 UTC), reflecting link quality variations caused by satellite mobility. While the ``Lookup-Table'' method also identifies the V2A workflow as optimal through historical querying, its execution remains inflexible. As shown, while the Lookup-Table-w/o Judg. variant achieves peak AKD performance, it does so at the cost of continuous, high-frequency knowledge base updates (totaling 17,284 symbols per segment). In contrast, the proposed adaptive strategy actively reduces the update level and reallocates bandwidth to enhance semantic protection. This strategy achieves performance comparable to Lookup-Table-w/o Judg. while consuming approximately 50\% less bandwidth. Meanwhile, Lookup-Table-L2, with a similar bandwidth allocation, consistently performs worse due to its lack of environmental understanding and dynamic planning capabilities.

\CheckRmv{\begin{figure} 
		\centerline{\includegraphics[width=3.0in]{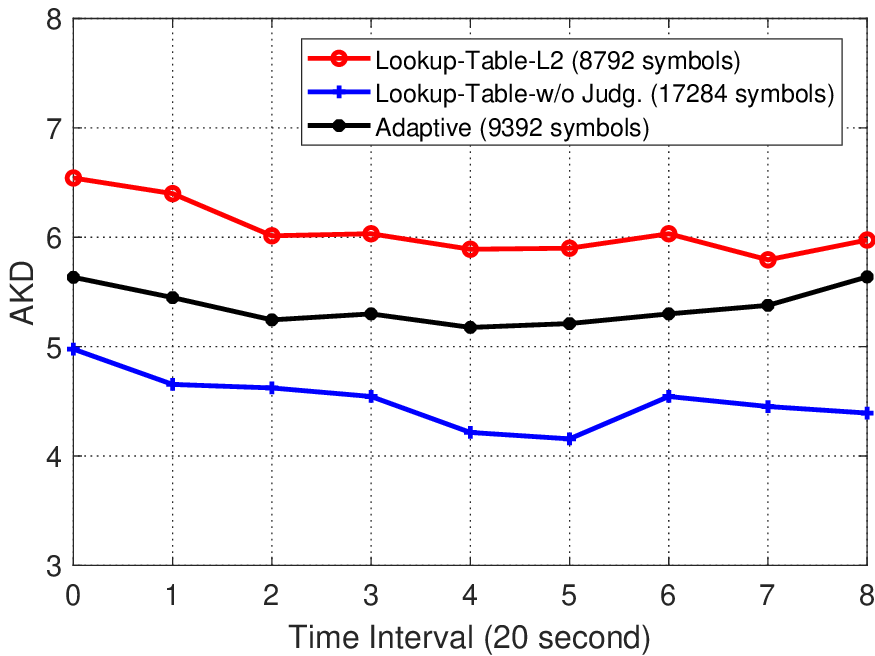}}
		\caption{Performance of different strategies during the service time.}
		\label{llm}
\end{figure}}

In summary, this case study demonstrates that the proposed adaptive decision-making framework deeply integrates environmental awareness, task understanding, and resource planning. Through system-level resource coordination and optimization, it effectively balances semantic fidelity and transmission overhead under stringent bandwidth constraints, significantly enhancing the overall robustness and resource efficiency of the system.

\section{Conclusion} \label{section:Conclusion}
This paper proposes an adaptive multimodal semantic transmission system for satellite communications, addressing the limitation of rigid transmission rules under highly uncertain channel conditions. Unlike traditional methods that rely on fixed modal priorities, our system employs a dual-stream architecture capable of flexibly switching between video-driven and audio-driven generation paths, thereby ensuring task adaptability in complex scenarios. To optimize the trade-off between knowledge base maintenance costs and generation quality, a dynamic keyframe update mechanism is introduced. This mechanism maintains the knowledge base as needed, achieving a balance between generation quality and bandwidth consumption. Crucially, an intelligent decision module is developed by leveraging an LLM agent to orchestrate these components. By reasoning over wireless conditions and task requirements, this module replaces static heuristics with active transmission strategies, addressing complex decision-making problems in constrained environments. Simulation results demonstrate the superiority of integrating cross-modal generation with intelligent management in satellite links. This architecture provides a robust foundation for next-generation multimedia satellite networks, enabling high-fidelity communication in resource-limited environments.

\bibliographystyle{IEEEtran}
\bibliography{my_ref}

\vfill
	
\end{document}